# Exceptionally Long-ranged Lattice Relaxation in Oxygen-deficient $Ta_2O_5$

Yong Yang[1,2], Osamu Sugino[1,3], Yoshiyuki Kawazoe[4,5]

[1]*National Institute for Materials Science (NIMS), 1-2-1 Sengen, Tsukuba, Ibaraki 305-0047, Japan*

[2]*Key Laboratory of Materials Physics, Institute of Solid State Physics, Chinese Academy of Sciences, Hefei 230031, China*

[3]*The Institute for Solid State Physics (ISSP), the University of Tokyo, 5-1-5 Kashiwanoha, Kashiwa, Chiba 277-8581, Japan*

[4]*New Industry Creation Hatchery Center (NICHe), Tohoku University, 6-6-4 Aoba, Aramaki, Aoba-ku, Sendai, Miyagi 980-8579, Japan*

[5]*Institute of Thermophysics, Siberian Branch of Russian Academy of Sciences, Novosibirsk 630090, Russia.*

The lattice relaxation in oxygen-deficient $Ta_2O_5$ is investigated using first-principles calculations. The presence of a charge-neutral oxygen vacancy can result in a long-ranged lattice relaxation which extends beyond 18 Å from the vacancy site. The lattice relaxation has significant effects on the vacancy formation energy as well as the electronic structures. The long-ranged behavior of the lattice relaxation is explained in terms of the Hellmann-Feynman forces and the potential energy surface related to the variation of Ta-O bond lengths.

Subject Areas: Computational Physics, Condensed Matter Physics, Materials Science

# I. INTRODUCTION

Metal oxides are a group of materials which play a significant role in modern physics, chemistry and technology [1-17]. The presence of defects (intrinsic or extrinsic) can result in nontrivial modifications on the electronic properties of metal oxides. In fact, the emergence of high-$T_c$ superconductivity or colossal magneto-resistance is due to the doping of transition metal oxides or rare earth metal oxides [2, 18-21]. The "F-centers" in MgO are due to the oxygen vacancies [22-24]. Presence of native defects can induce magnetism in a number of metal oxides including $TiO_2$ [25, 26], $V_2O_5$ [27], $HfO_2$ [28], $BaTiO_3$ [29], CaO [30] and ZnO [31], whose defect-free structures are nonmagnetic. Oxygen vacancies also play a key role in the leakage current in $Ta_2O_5$ [13, 32], a material of importance to both dielectrics [13, 32, 33] and catalysis applications [34, 35]. With the presence of oxygen vacancies, the functionality of electrode materials [36] can be improved, and the catalytic behavior of metal oxides can be significantly enhanced [37, 38]. Band gap narrowing effects are reported in anatase $TiO_2$ by the co-doping of extrinsic defects [39].

Numerous works have been devoted to studying the defects in metal oxides. In principle, the structural properties of defects on surfaces can be characterized by experimental tools such as scanning tunneling microscope (STM) [37, 38, 40, 41] or scanning electron microscope (SEM) [42], and defects in bulk by the transmission electron microscope (TEM) [43] or X-ray diffraction (XRD) [44, 45]. The electronic structures of defects can be detected from the measurements including electrical transport and optical absorption [46, 47], the electron paramagnetic resonance (EPR) [48-51] and the X-ray photoemission spectroscopy (XPS) [52-55]. In addition to the experimental methods, first-principles calculations can provide insightful details such as charge transfer, vacancy formation energy, and bonding geometry of the defects, which are not easily accessed by the state-of-the-art experimental tools.

In the simulation of bulk or surfaces defects of solids, the so-called supercell approach is widely employed, by which the fundamental equation of quantum mechanics is solved within a unit cell containing a number of atoms subject to the periodic boundary condition [56, 57]. The success of supercell approach is due to the fact that the defect-induced lattice distortion is generally short-ranged especially for a neutral defect. This is, however *not* always true as has been shown by hybrid classical-*ab initio* embedded-cluster calculations for a neutral vacancy in α-quartz [58]: The lattice displacement of the order 0.05 Å extends over 13 Å from the vacant site, although such long-ranged behavior has not been reported from first-principles supercell calculation being possibly obscured by the limited size of the simulation cell, which is typically ~ 10Å × 10Å × 10Å [59-67]. It is natural to consider that similar long-ranged relaxation may be found likewise for a material having large dielectric constant and soft phonon mode, in which the perturbation by a defect is strong and can easily extend.

In this context, we focus on $Ta_2O_5$, a candidate material which has a large dielectric constant [13, 68-75] and a variety of coexisting polymorphs that tend to stabilize local metastable structures [76]. In doing the research, instead of the powerful hybrid classical-*ab initio* method [58], the supercell approach can be alternatively utilized to capture the long-ranged nature of the lattice relaxation when the supercell is large enough. We investigate the lattice relaxation in a direction using a supercell which sufficiently elongates in the corresponding unit cell axis (~ 41 Å). From the calculation we find that the lattice relaxation associated with charge-neutral vacancy can extend beyond 18 Å from the vacant site. The lattice relaxation has significant effects on the vacancy formation energy as well as the electronic structures. The result is analyzed using the long-ranged Hellmann-Feynman forces and the density response of electrons as well as the potential energy surface which indicates softness of the Ta-O bonds.

## II. METHODS OF INVESTIGATION

Our study is based on first-principles calculations carried out by the Vienna *ab initio* simulation package (VASP) [77, 78], using a plane wave basis set and the projector-augmented-wave (PAW) potentials [79, 80]. For structural relaxation, the exchange-correlation interactions of electrons are described by the PBE functional [81]. To overcome the problem of band gap underestimation, the PBE0 hybrid functional [82] is employed for the calculation of electronic density of states. The energy cutoff for plane waves is 600 eV. The construction of the primitive cell of $Ta_2O_5$ (low temperature phase) is detailed in our recent work [83]. For the simulation of an oxygen vacancy, we take a (2×1×2) supercell of $Ta_2O_5$, whose geometric parameters are as follows: $a$ = 12.664 Å, $b$ = 40.921 Å, $c$ = 7.692 Å; $α$ = 90º, $β$ = 90º, $γ$ = 89.16º. A 1×1×2 k-mesh generated using the Monkhorst-Pack scheme [84] is employed for all the calculations.

As discussed in previous works [83, 85], the constituent O atoms in $Ta_2O_5$ can be classified according to their geometric positions: O on the basal planes (referred to as in-plane site) and O sitting between the basal planes (referred to as cap sites). The properties of an oxygen vacancy are studied by removing one O atom in the supercell of $Ta_2O_5$. We have performed calculations on a number of vacancy structures whose geometries and energetic parameters can be found elsewhere [83]. Here, we will focus on two typical vacancy configurations: The *cap site* vacancy, i.e., the vacancy configuration A in Ref. [83] (referred to as Vo I hereafter), and the *in-plane site* vacancy (vacancy configuration D in Ref. [83], referred to as Vo II). We will study the lattice relaxation induced by the charge-neutral state of these two vacancy configurations. Due to the application of periodic boundary conditions in the calculations, the single-vacancy configurations that we study in this work are actually periodically arranged oxygen vacancies with a dilute vacancy concentration of ~ 0.45%. This can be regarded as an approximation for modeling a point defect.

### III. LATTICE RELAXATION AROUND THE VACANCIES

The lattice relaxation can be characterized by two ways: 1) Variation of the Ta-O bond lengths and 2) displacement from the ideal positions. We take the first way for the sake of simplicity since the second way requires us to examine the displacement of both Ta and O atoms separately. Figure 1 shows the variation of Ta-O bond lengths plotted against the distance R from the ideal vacancy (Vo) site. For each distance R, plotted are the maximum, minimum and average of the variation. The lattice relaxation is strikingly long-ranged for the vacancy at the cap site (Vo I) and is much shorter for the vacancy at the inplane site (Vo II). For Vo I, the maximum of the absolute deviation $|\delta R_{Ta-O}|$ remains 0.025 Å even when the distance R increases to 20 Å. Interestingly, compared with $|\delta R_{Ta-O}|$ near the vacancy sites (R ≤ ~ 3 Å), larger values can be found at longer distances (~ 6 Å ≤ R ≤ ~ 10 Å). For both Vo I and Vo II, though $|\delta R_{Ta-O}|$ shows a general trend of decreasing with increasing R, the distribution of $|\delta R_{Ta-O}|$ is highly fluctuated, which is in direct contrast to the intuitive experience from a continuum model, in which one would expect that the distortion decays monotonically with the distance to the vacancy site. For distances beyond 10 Å, the lattice relaxation extends mainly along the *b*-axis. In this context, a larger sized simulation unit cell, such as a (2×1×3) or (3×1×3) supercell of $Ta_2O_5$ may be necessary to evaluate the relaxation effects of vacancy in both *a*-axis and *c*-axis. Such simulations are beyond the affordable computational resources at present stage and are left for the future study.

Figure 2 shows the Hellmann-Feynman (HF) forces [86] acting on the atoms of the *unrelaxed* defective structures. The distance within which the HF forces are not negligible is ~ 19 Å for Vo I and is ~ 9 Å for Vo II. The forces are long-ranged when compared with other systems such as oxygen-deficient β-$PtO_2$ [87] (see also the figure in the Appendix), in which the bond length distortion and the HF forces are negligible for the distances beyond 6 Å from the vacancy site. Compared to the variation of the Ta-O bond lengths, the magnitude of HF forces is less fluctuated and has the largest value near the vacancy site, R ~ 2.5 Å. One can see the correspondence of the forces and bond length variation from Figs. 1-2 although it is not accurately one-to-one, e.g., the global maximum of HF forces does not correspond to the global maximum of $|\delta R_{Ta-O}|$. The imperfect correspondence is understandable when taking into account the fact that the lattice relaxation is not the behavior of individual atom but the collective motions of all the atoms. Yet, for both Vo I and Vo II, in the region of R where the value of HF force is scattered the bond length variation $|\delta R_{Ta-O}|$ is scattered as well. Finding a quantitative relation between the bond length variation and the HF forces will be an important problem but will be left for a future study.

We go on to study the effects of lattice relaxation on the vacancy formation energy and the electron density of states (DOS). For both Vo I and Vo II, three structures are considered: the unrelaxed structure; the partially relaxed structure, in which only the atoms located in a distance of R ≤ 6 Å to the Vo site are allowed to relax; the fully relaxed structure, where all the atoms are relaxed. The calculated vacancy formation energies are summarized in Table I. The differences in the vacancy formation energy reflect the significant role of lattice relaxation. It is emphasized that the restricted lattice relaxation allowed for the partially relaxed structure is so insufficient that the relaxation energy is increased by ~ 0.67 eV for Vo I and ~ 0.36 eV for Vo II. Under the conditions of thermodynamic equilibrium in which the defective systems are subjected to *high temperatures* [88, 89], Vo II would show higher stability than Vo I because of its lower formation energy. In other words, Vo I and the other Vo configurations can be stabilized at *low temperatures* where the diffusion of vacancies is significantly hindered by the low defect kinetics and high diffusion barrier (the order of vacancy formation energy). The DOS analysis is focused on the electronic states near the valence band maximum (VBM) and the conduction band minimum (CBM). The calculated DOS using PBE0 functional is shown in Fig. 3: Left panels for Vo I and right panels for Vo II. For the three differently relaxed structures of Vo I, clear differences are found at the following aspects: The number of gap states below the Fermi level ($E_F$), the position of the gap states, and the position of $E_F$. In the case of Vo II, the DOS features of the three structures differ slightly around the VBM and CBM, e.g., the peak position of the gap state near CBM (marked in Fig. 3). On the other hand, the position of $E_F$, and the overall DOS of the partially-relaxed and fully-relaxed structure are similar, which is in accordance with the relatively small difference in the vacancy formation energy of both structures of Vo II (Table I).

Then, what is the origin of the long-ranged HF forces? Similar to molecules or atomic clusters [90], the HF forces on the atoms of a crystal (regarded as a huge molecule) may be expressed as [90, 91],

$$\vec{F}_{R_I} = \int n(\vec{r}) \nabla V(\vec{r} - \vec{R}_I) d\vec{r} + \vec{F}_{nuc} = -Z_I \int n(\vec{r}) \frac{\vec{r} - \vec{R}_I}{|\vec{r} - \vec{R}_I|^3} d\vec{r} + \sum_{J, J \neq I}^{N} Z_I Z_J \frac{\vec{R}_J - \vec{R}_I}{|\vec{R}_J - \vec{R}_I|^3}, \quad (1)$$

where $n(\vec{r})$ is the electron density, $V(\vec{r} - \vec{R}_I) = -Z_I / (\vec{r} - \vec{R}_I)$ is the electron-nucleus interaction and $Z_I$, $Z_J$ are the effective charge numbers of the nucleus (or the ion core in the case of DFT pseudopotential calculations). The total number of atoms is $N$. The coordinates of electrons and atoms are $\vec{r}$ and $\vec{R}_I$ ($I = 1, \ldots, N$), respectively. The term $\vec{F}_{nuc} = \sum_{J, J \neq I}^{N} Z_I Z_J \frac{\vec{R}_J - \vec{R}_I}{|\vec{R}_J - \vec{R}_I|^3}$ is the electrostatic force on the atom *I* due to the presence of other nuclei/ion cores. For a perfect crystal or relaxed atomic structure, the relation $\vec{F}_{R_I} = 0$ holds for each atom. Compared with the perfect crystal, the two *unrelaxed* defective structures discussed in Fig. 2 correspond to the situation in which the removal of an oxygen atom causes changes in the electron density $n(\vec{r})$ and the term $\vec{F}_{nuc}$, i.e., $n(\vec{r}) \rightarrow n(\vec{r}) + \delta n(\vec{r})$ and $\vec{F}_{nuc} \rightarrow \sum_{J, J \neq I}^{N} Z_I Z_J \frac{\vec{R}_J - \vec{R}_I}{|\vec{R}_J - \vec{R}_I|^3} - Z_I Z_{Vo} \frac{\vec{R}_{Vo} - \vec{R}_I}{|\vec{R}_{Vo} - \vec{R}_I|^3}$. It is straight forward that the HF forces in the *unrelaxed* defective structures are as follows:

$$\vec{f}_{R_I} = -Z_I \int \delta n(\vec{r}) \frac{\vec{r} - \vec{R}_I}{|\vec{r} - \vec{R}_I|^3} d\vec{r} - Z_I Z_{Vo} \frac{\vec{R}_{Vo} - \vec{R}_I}{|\vec{R}_{Vo} - \vec{R}_I|^3}. \tag{2}$$

The change of electron density $\delta n(\vec{r})$ consists of two parts: the electron density of the removed O atom in the bulk phase, $n_O(\vec{r})$, and the charge density perturbation due to the removal of the O atom, $\Delta n(\vec{r})$, i.e., $\delta n(\vec{r}) = n_O(\vec{r}) + \Delta n(\vec{r})$. The term $n_O(\vec{r})$ is localized around the Vo site and satisfies the relation $\int n_O(\vec{r}) d\vec{r} = -Z_{Vo}$, with the induced forces equal to $-Z_I \int n_O(\vec{r}) \frac{\vec{r} - \vec{R}_I}{|\vec{r} - \vec{R}_I|^3} d\vec{r}$, which will largely cancel the second term in Eq. (2). For the atoms which locate far away from the Vo site, only the charge density perturbation $\Delta n(\vec{r})$ plays an important role in inducing the HF forces. Long-ranged density perturbation leads to long-ranged HF forces.

To demonstrate the effects of perturbation on charge density upon the creation of an oxygen vacancy, we compute the electron density difference between the perfect $Ta_2O_5$ (reference system) and the *unrelaxed* Vo configuration, for Vo I and Vo II, plus one O atom at the corresponding Vo site. The difference is given by $\Delta n = n[Ta_2O_5] - (n[Vo] + n[O])$, where $n[Vo]$ and $n[Ta_2O_5]$ are the total electron density of the system with and without an oxygen vacancy, respectively. The term *n*[O] is the electron density of an isolated O atom. The results are shown in Fig. 4. The perturbation is large near the oxygen vacancy site due to the breaking of the Ta-O bonds, and then decreases with increasing distance [Figs. 4 (a)-(b)]. The charge perturbation is relatively large and extended in Vo I than in Vo II. In the long distance, the perturbation in Vo I is small but extends in the whole simulation cell without showing an obvious decay. For both cases, Vo I and Vo II, the range of electron density perturbation is consistent with that of the HF forces. Considering the fact that the positions of the atoms of the *unrelaxed* defective structures are exactly the same as the corresponding ones in the perfect structures, the HF forces in regions far away from the Vo site are triggered mainly by the electron density perturbation.

Another factor that may be responsible for the long-ranged lattice relaxation is the potential energy surface that determines the force constants for the variation of the Ta-O bond lengths. Soft force constants imply flexible bond lengths. We illustrate softness of the force constant by varying the lengths of a certain Ta-O bond around the optimized one. In the variation, the positions of the Ta and O atoms which form the Ta-O bond are fixed while the positions of the other atoms are either fixed/unrelaxed or relaxed to show the effects of lattice relaxation on total energy. Figure 5 shows the associate increase in the total energy for the unrelaxed structures. Here those bonds that show large variation (Bonds 1, 2, and 3 marked in Fig. 1) were chosen for the analysis. There are flat regions at which the change in total energy is remarkably small especially for Bond 2. When the other atoms are relaxed, the increase in the total energy is much smaller, as illustrated in Fig. 6 for Bond 1 that shows the largest variation (Fig. 1). Interestingly, for Vo II, we find a new global minimum structure at which the bond length is $R_{TaO} \sim 1.9$ Å, whose total energy is slightly lower than the previous one ($R_{TaO} \sim 2.9$ Å). The finding of the new global minimum also implies that the geometry optimization of our previous study is incomplete and imposes more systematic search for the global minimum. However, searching for the global minimum remains a challenge in computational condensed matter physics, due to the complexity of high-dimensional potential energy surface. The above results illustrate softness of the lattice which can explain the appearance of long-ranged lattice relaxation. The soft lattice might be related to coexistence of metastable polymorphs in $Ta_2O_5$. Quantitative investigation of the relation is beyond the scope of this paper and is subject of future study.

In summary, we have shown that the presence of charge-neutral oxygen vacancies in $Ta_2O_5$ is associated with exceptionally long-ranged lattice relaxation, which has notable effects on both the vacancy formation energy and the vacancy states of electrons. The results are explained by the long-ranged Hellmann-Feynman forces and the soft force constants for the variation of Ta-O bond lengths. The long-ranged character of forces is mainly caused the long-ranged perturbation of charge density, which is peculiar to this system. The smooth potential energy surface of the relaxed structures also indicates that the ground state of $Ta_2O_5$ crystal structure (with and without oxygen vacancy) is highly degenerate, in accordance with the coexistence of polymorphs and the inherent amorphous nature of experimentally grown $Ta_2O_5$ [74,

92].

**Acknowledgments:** This work is partly supported by the Global Research Center for Environment and Energy based on Nanomaterials Science (GREEN) at National Institute for Materials Science. The first-principles simulations are performed using the supercomputers of NIMS, the supercomputers of the Hefei Branch of Supercomputing Center of Chinese Academy of Sciences, and the supercomputers of the Institute for Materials Research, Tohoku University.

**Appendix:**

In this appendix, we show that the lattice relaxation and HF forces in oxygen-deficient β-PtO$_2$ [87] are short-range:

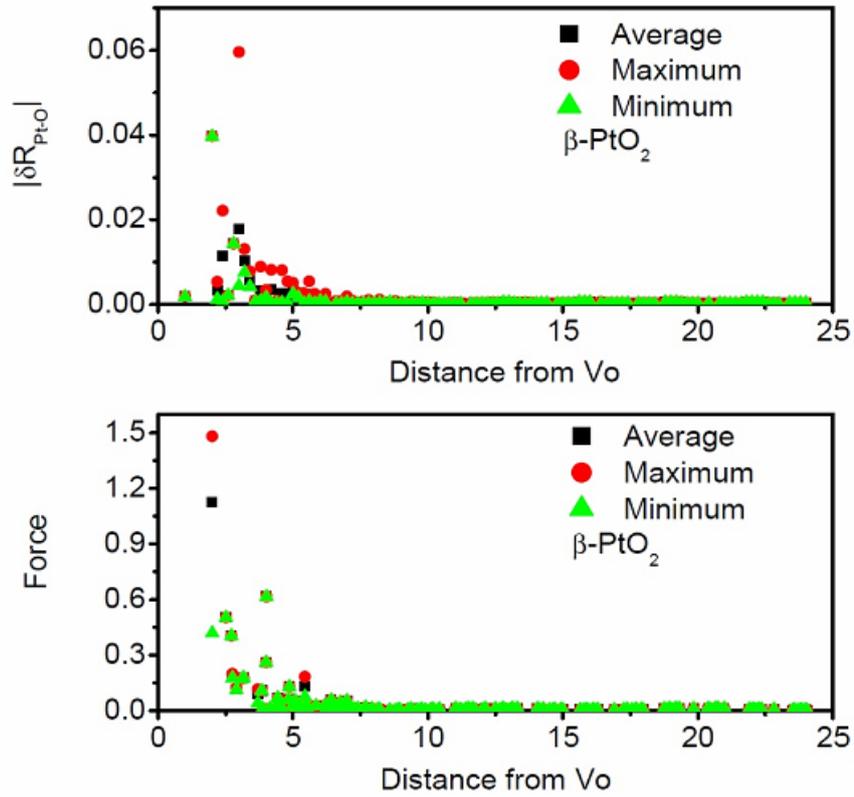

**FIG. A1** The maximum, minimum and average absolute variation of Pt-O bond lengths, and the HF forces on the atoms (unrelaxed structure) in oxygen-deficient β-PtO$_2$, as a function of distance to the Vo site. The unit for lengths is Å, and for forces is eV/Å.

**References:**

[1] J. G. Bednorz, K. A. Mueller, *Z. Phys. B* **64**, 189 (1986).

[2] M. K. Wu, J. R. Ashburn, C. J. Torng, P. H. Hor, R. L. Meng, L. Gao, Z. J. Huang, Y. Q. Wang, and C. W. Chu, *Phys. Rev. Lett.* **58**, 908 (1987).

[3] N. F. Mott, *Proceedings of the Physical Society*. Series A **62**, 416 (1949).

[4] P. Kuiper, G. Gruizinga, J. Ghijsen, G. A. Sawatzky, H. Verweij, *Phys. Rev. Lett.* **62**, 221 (1987).

[5] Y. Tokura and N. Nagaosa, *Science* **288**, 462 (2000).

[6] A. Fujishima and K. Honda, *Nature* (London) **238**, 37 (1972).

[7] A. Fujishima, T. N. Rao, D. A. Tryk, *J. PhotoChem. & PhotoBio. C: PhotoChem. Rev.* **1**, 1 (2000).

[8] M. R. Hoffmann, S. T. Martin, W. Choi, and D. W. Bahnemannt, *Chem. Rev.* **95**, 69 (1995).

[9] A. L. Linsebigler, G. Lu, and J. T. Yates, Jr., *Chem. Rev.* **95**, 735 (1995).

[10] "Doping and Functionalization of Photoactive Semiconducting Metal Oxides", special issue edited by C. Di Valentin, U. Diebold, and A. Selloni [*Chem. Phys.* **339**, 1-129 (2007)].

[11] P. Poizot, S. Laruelle, S. Grugeon, L. Dupont & J-M. Tarascon, *Nature* (London) **407**, 496 (2000).

[12] R. A. Miller, *J. Thermal Spray Tech*. **6**, 35 (1997).

[13] H. Kimura, J. Mizuki, S. Kamiyama and H. Suzuki, *Appl. Phys. Lett.* **66**, 2209 (1995).

[14] S. Sayan, R. A. Bartynski, X. Zhao, E. P. Gusev, D. Vanderbilt, M. Croft, M. Banaszak Holl, and E. Garfunkel, *Phys. Stat. Sol.* (b) **241**, 2246 (2004).

[15] M. V. Ganduglia-Pirovano, A. Hofmann, J. Sauer, *Surf. Sci. Rep.* **62,** 219 (2007).

[16] C. Ronning, P. X. Gao, Y. Ding, Z. L. Wang, D. Schwen, *Appl. Phys. Lett.* **84**, 783 (2004).

[17] S. J. Pearton, D. P. Norton, Y. W. Heo, L. C. Tien, M. P. Ivill, Y. Li, B. S. Kang, F. Ren, J. Kelly, and A. F. Hebard, *J. Electron Materials* **35**, 862 (2006).

[18] Y. Kamihara, *et al.*, *J. Am. Chem. Soc.* **130**, 3296 (2008).

[19] X. H. Chen, *et al.*, *Nature* (London) **453**, 761 (2008).

[20] Z. A. Ren, *et al.*, *Europhys. Lett.* **83**, 17002 (2008).

[21] S. Jin, T. H. Tiefel, M. McCormack, R. A. Fastnacht, R. Ramesh, L. H. Chen, *Science* **264**, 413 (1994).

[22] Q. S. Wang and N. A. W. Holzwarth, *Phys. Rev. B* **41**, 3211 (1990).

[23] G. Pacchioni, *Solid State Sciences* **2**, 161 (2000).

[24] P. Mori-Sánchez, *et al.*, *Phys. Rev. B* **66**, 075103 (2002).

[25] S. Zhou, *et al.*, *Phys. Rev. B* **79**, 113201 (2009).

[26] K. Yang, *et al.*, *Phys. Rev. B* **81**, 033202 (2010).

[27] Z. R. Xiao and G. Y. Guo, *J. Chem. Phys.* **130**, 214704 (2009).

[28] C. D. Pemmaraju and S. Sanvito, *Phys. Rev. Lett.* **94**, 217205 (2005).

[29] D. Cao, M. Q. Cai, Y. Zheng and W. Y. Hua, *Phys. Chem. Chem. Phys.* **11**, 10934 (2009).

[30] I. S. Elfimov, S. Yunoki and G. A. Sawatzky, *Phys. Rev. Lett.* **89**, 216403 (2002).

[31] Q. Wang, Q. Sun, G. Chen, Y. Kawazoe and P. Jena, *Phys. Rev. B* **77**, 205411 (2008).

[32] R. M. Fleming, *et al.*, *J. Appl. Phys.* **88**, 850 (2000).

[33] S. Clima, *et al.*, *J. Electrochem. Soc.* **157**, G20 (2010).

[34] Y. Zhu, F. Yu, Y. Man, Q. Tian, Y. He, and N. Wu, *J. Solid State Chem.* **178**, 224 (2005).

[35] A. Ishihara, M. Tamura, K. Matsuzawa, S. Mitsushima, and K. Ota, *Electrochimica Acta* **55**, 7581 (2010).

[36] W. Y. Ma, B. Zhou, J. F. Wang, X. D. Zhang and Z. Y. Jiang, *J. Phys. D: Appl. Phys.* **46**, 105306 (2013).

[37] R. Schaub, *et al.*, *Phys. Rev. Lett.* **87**, 266104 (2001).

[38] E. Wahlström, *et al.*, *Phys. Rev. Lett.* **90**, 026101 (2003).

[39] W. Zhu., *et al.*, *Phys. Rev. Lett.* **103**, 226401 (2009).

[40] R. Schaub, *et al.*, *Science* **299**, 377 (2003).

[41] Martin Setvín *et al.*, *Science* **341**, 988 (2013).

[42] J. Goldstein, D. E. Newbury, D. C. Joy, C. E. Lyman, P. Echlin, E. Lifshin, L. Sawyer, and J. R. Michael. *Scanning Electron Microscopy and X-ray Microanalysis* (3 ed). Springer, 2003.

[43] T. Metzger, *et al.*, *Phil. Mag. A* **77**, 1013 (1998).

[44] H. Heinke, V. Kirchner, S. Einfeldt, and D. Hommel, *Appl. Phys. Lett.* **77**, 2145 (2000).

[45] K. Ueda, H. Tabata, and T. Kawai, *Appl. Phys. Lett.* **79**, 988 (2001).

[46] H. Tang, K. Prasad, R. Sanjinès, P. E. Schmid, and F. Lévy, *J. Appl. Phys.* **75**, 2042 (1994).

[47] P. Kofstad, *Oxidation of Metals* **44**, 3 (1995).

[48] P. M. Lenahan and P. V. Dressendorfer, *J. Appl. Phys.* **55**, 3495 (1984).

[49] J. A. Weil, *Phys. Chem. Minerals* **10**, 149 (1984).

[50] N. G. Kakazey, *et al.*, *J. Mater. Sci.* **32**, 4619 (1997).

[51] A. Stesmans, V. V. Afanas'ev and K. Clémer, *ECS Transactions* **1**, 347 (2006).

[52] S. Uhlenbrock, C. Scharfschwerdt, M. Neumann, G. Illing and H.-J. Freund, *J. Phys.: Condens. Matter* **4**, 7973 (1992).

[53] M. Batzill, E. H. Morales, and U. Diebold, *Phys. Rev. Lett.* **96**, 026103 (2006).

[54] H. B. Fan, *et al.*, *Chin. Phys. Lett.* **24**, 2108 (2007).

[55] M. Forker, *et al.*, *Phys. Rev. B* **77**, 054108 (2008).

[56] M. C. Payne, M. P. Teter, D. C. Allan, T. A. Arias and J. D. Joannopoulos, *Rev. Mod. Phys.* **64**, 1045 (1992).

[57] C. G. Van de Walle and J. Neugebauer, *J. Appl. Phys.* **95**, 3851 (2004).

[58] V. B. Sulimov, P. V. Sushko, A. H. Edwards, A. L. Shluger, and A. Marshall Stoneham, *Phys. Rev. B* **66**, 024108 (2002).

[59] A. F. Kohan, G. Ceder, D. Morgan, C. G. Van de Walle, *Phys. Rev. B* **61**, 15019 (20010).

[60] A. S. Foster, V. B. Sulimov, F. Lopez Gejo, A. L. Shluger, and R. M. Nieminen, *Phys. Rev. B* **64**, 224108 (2001).

[61] K. Hermann, M. Witko, R. Druzinic, R. Tokarz, *Appl. Phys. A* **72**, 429 (2001).

[62] A. S. Foster, V. B. Sulimov, F. Lopez Gejo, A. L. Shluger, and R. M. Nieminen, *J. Non-Cryst. Solids* **303**, 101 (2002).

[63] J. Carrasco, N. Lopez, and F. Illas, *Phys. Rev. Lett.* **93**, 225502 (2004).

[64] J. Carrasco, N. Lopez, and F. Illas, *J. Chem. Phys.* **122**, 224705 (2005).

[65] P. Erhart, K. Albe, and A. Klein, *Phys. Rev. B* **73**, 205203 (2006).

[66] A. M. Ferrari, C. Pisani, F. Cinquini, L. Giordano, and G. Pacchioni, *J. Chem. Phys.* **127**, 174711 (2007).

[67] D. O. Scanlon, A. Walsh, B. J. Morgan, and G. W. Watson, *J. Phys. Chem. C*, **112**, 9903 (2008).

[68] G. S. Oehrlein, F. M. d'Heurle, and A. Reisman, *J. Appl. Phys.* **55**, 3715 (1984).

[69] A. Pignolet, G. M. Rao, and S. B. Krupanidhi, *Thin Solid Films* **258**, 230 (1995).

[70] R. J. Cava, W. F. Peck, and J. J. Krajewski, *Nature* **377**, 215 (1995).

[71] R. J. Cava, J. J. Krajewski, W. F. Peck, Jr., and G. L. Roberts, *J. Appl. Phys.* **80**, 2346 (1996).

[72] R. J. Cava and J. J. Krajewski, *J. Appl. Phys.* **83**, 1613 (1998).

[73] C. Chaneliere, S. Four, J. L. Autran, R. A. B. Devine, and N. P. Sandler, *J. Appl. Phys.* **83**, 4823 (1998).

[74] P. C. Joshi and M. W. Cole, *J. Appl. Phys.* **86**, 871 (1999).

[75] J. Lin, N. Masaaki, A. Tsukune, and M. Yamada, *Appl. Phys. Lett.* **74**, 2370 (1999).

[76] C. Askeljung, B.-O. Marinder, and M. Sundberg, *J. Solid State Chem.* **176**, 250 (2003).

[77] G. Kresse and J. Hafner, *Phys. Rev. B* **47**, 558 (1993).

[78] G. Kresse and J. Furthmüller, *Phys. Rev. B* **54**, 11169 (1996).

[79] P. E. Blöchl, *Phys. Rev. B* **50**, 17953 (1994).

[80] G. Kresse and D. Joubert, *Phys. Rev. B* **59**, 1758 (1999).

[81] J. P. Perdew, K. Burke, M. Ernzerhof, *Phys. Rev. Lett.* **77**, 3865 (1996).

[82] J. P. Perdew, M. Ernzerhof, K. Burke, *J. Chem. Phys.* **105**, 9982 (1996).

[83] Y. Yang, H.-H. Nahm, O. Sugino, and T. Ohno, *AIP Advances* **3**, 042101 (2013).

[84] H. J. Monkhorst and J. D. Pack, *Phys. Rev. B* **13**, 5188 (1976).

[85] R. Ramprasad, *J. Appl. Phys.* **94**, 5609 (2003).

[86] R. P. Feynman, *Phys. Rev.* **56**, 340 (1939); R. M. Martin, "*Electronic Structure: Basic theory and practical methods*," Cambridge University Press, 2004.

[87] Y. Yang, O. Sugino and T. Ohno, *Phys. Rev. B* **85**, 035204 (2012).

[88] C. G. Van de Walle and J. Neugebauer, *J. Appl. Phys.* **95**, 3851 (2004).

[89] A. Glensk, B. Grabowski, T. Hickel, and J. Neugebauer, *Phys. Rev. X* **4**, 011018 (2014).

[90] http://en.wikipedia.org/wiki/Hellmann–Feynman_theorem.

[91] R. S. Sorbello and B. B. Dasgupta, *Phys. Rev. B* **21**, 2196 (1980).

[92] J. M. Ngaruiya, S. Venkataraj, R. Drese, O. Kappertz, T. P. Leervad Pedersen and M. Wuttig, *Phys. Status Solidi A* **198**, 99 (2003).

**TABLE I** Vacancy formation energy ($E_{vf}$) of differently relaxed structures of Vo I and Vo II.

| $E_{vf}$ (eV) | Unrelaxed | Partially Relaxed | Fully Relaxed |
|---|---|---|---|
| Vo I | 6.159 | 5.579 | 4.907 |
| Vo II | 5.317 | 4.528 | 4.164 |

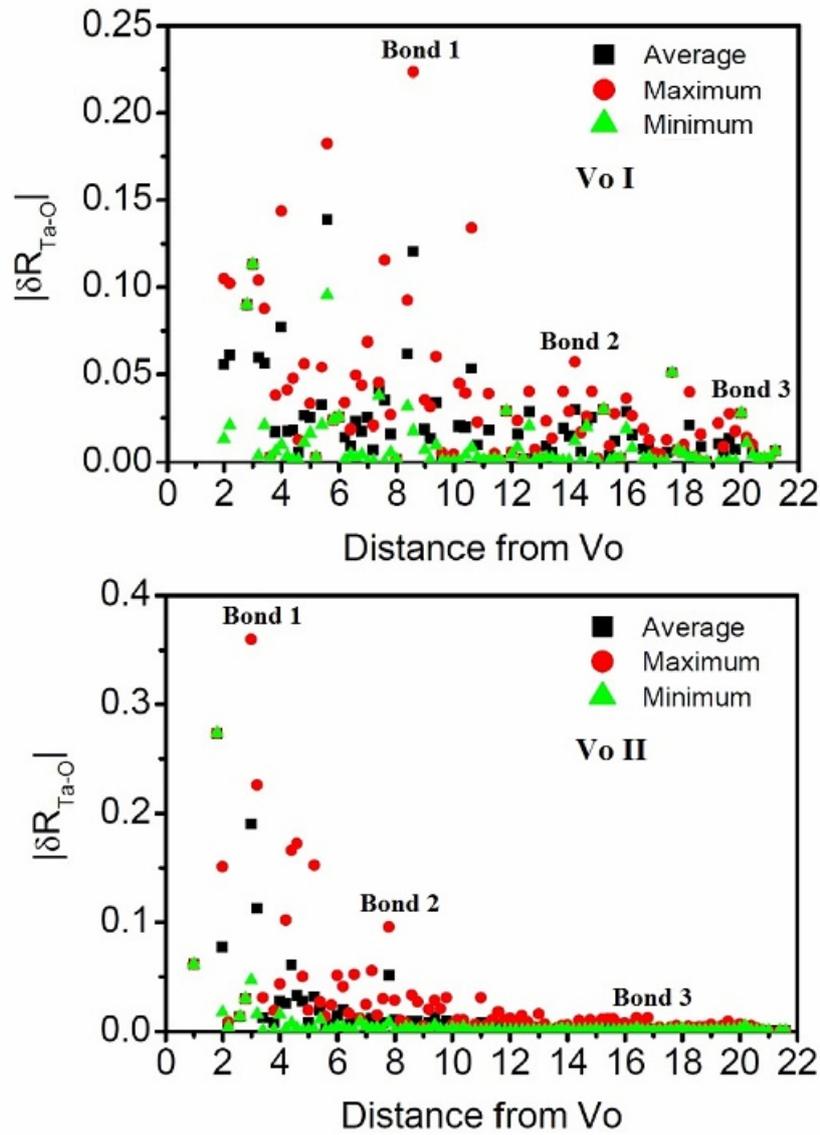

**FIG. 1** The maximum, minimum and average absolute deviation of Ta-O bond lengths (with referenced to the perfect structure) as a function of distance to the oxygen vacancy (Vo) site, for configurations Vo I and Vo II. The unit for lengths is Å.

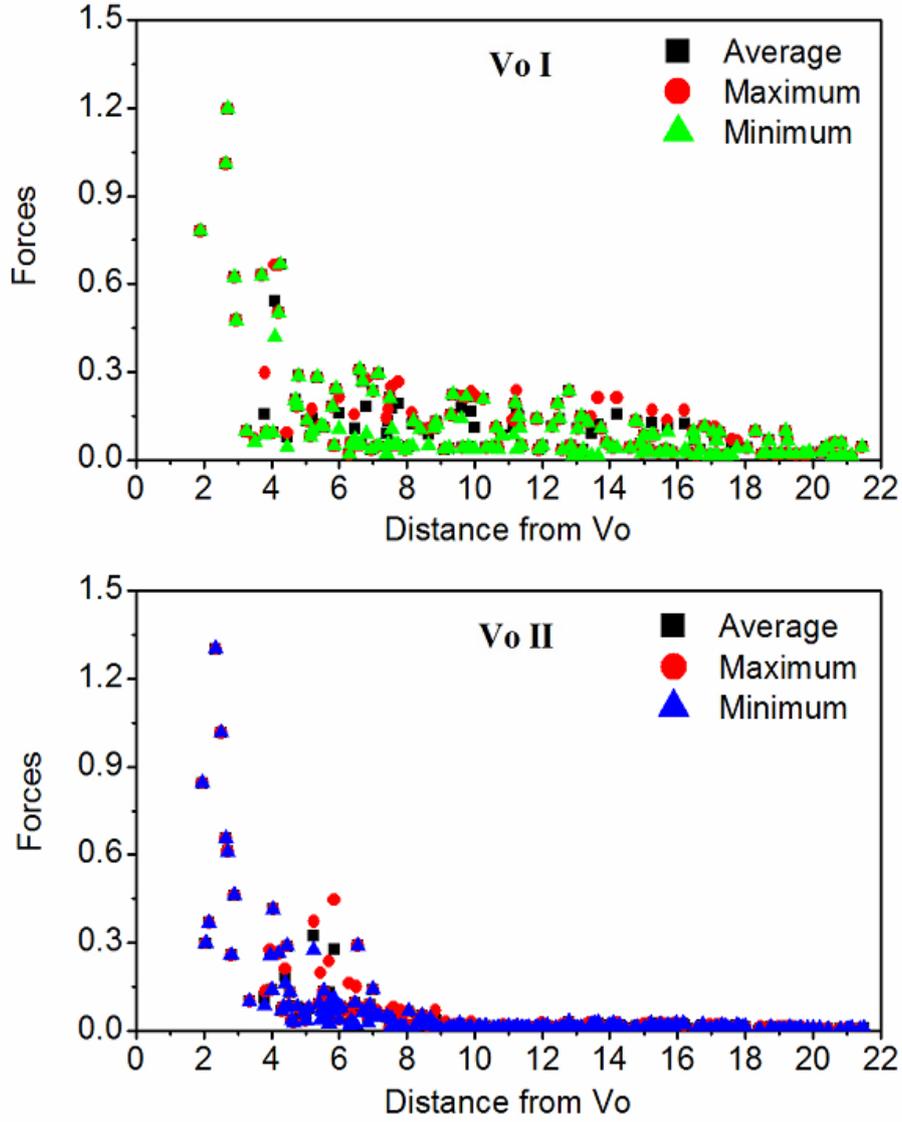

**FIG. 2** The maximum, minimum and average Hellmann-Feynman forces on the atoms, as a function of distance to the Vo site, for the unrelaxed configurations of Vo I and Vo II. The unit for lengths is Å, and for forces is eV/Å.

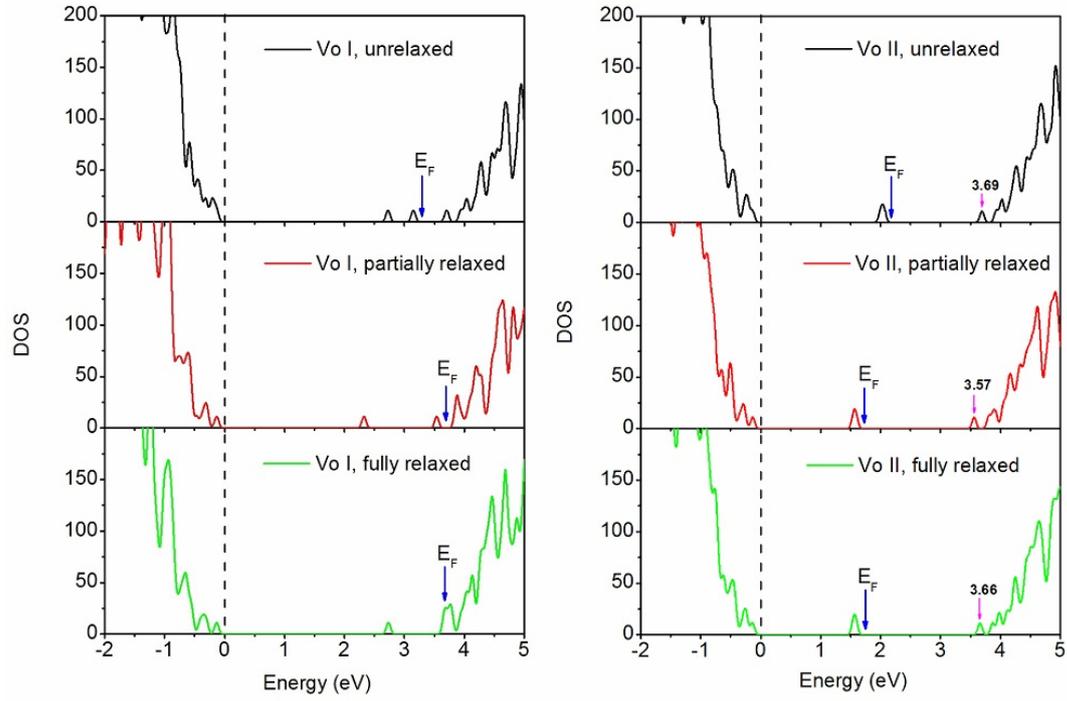

**FIG. 3** The electron DOS of unrelaxed, partially relaxed and fully relaxed structures of Vo I (Left Panels) and Vo II (Right Panels), calculated using the PBE0 functional [82]. The top of valence band is set at 0 (dashed lines), and the position of Fermi level ($E_F$) is marked by a vertical arrow. The unit for DOS is states/eV/supercell.

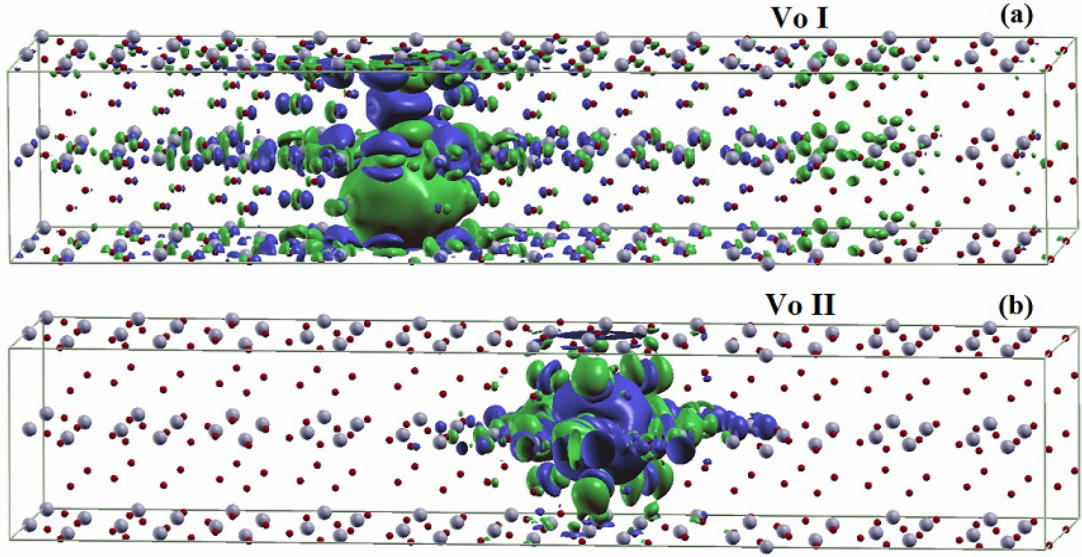

**FIG. 4 Panels (a)-(b)**: Electron density difference between the perfect structure of Ta$_2$O$_5$, and the unrelaxed Vo configuration plus one isolated O atom. For both systems (Vo I, Vo II), the isolated O atom is at spin singlet state, since the magnetic moment of oxygen atoms is zero in both perfect and defective Ta$_2$O$_5$. The isovalue for density is ± 0.002 $e$/Å$^3$.

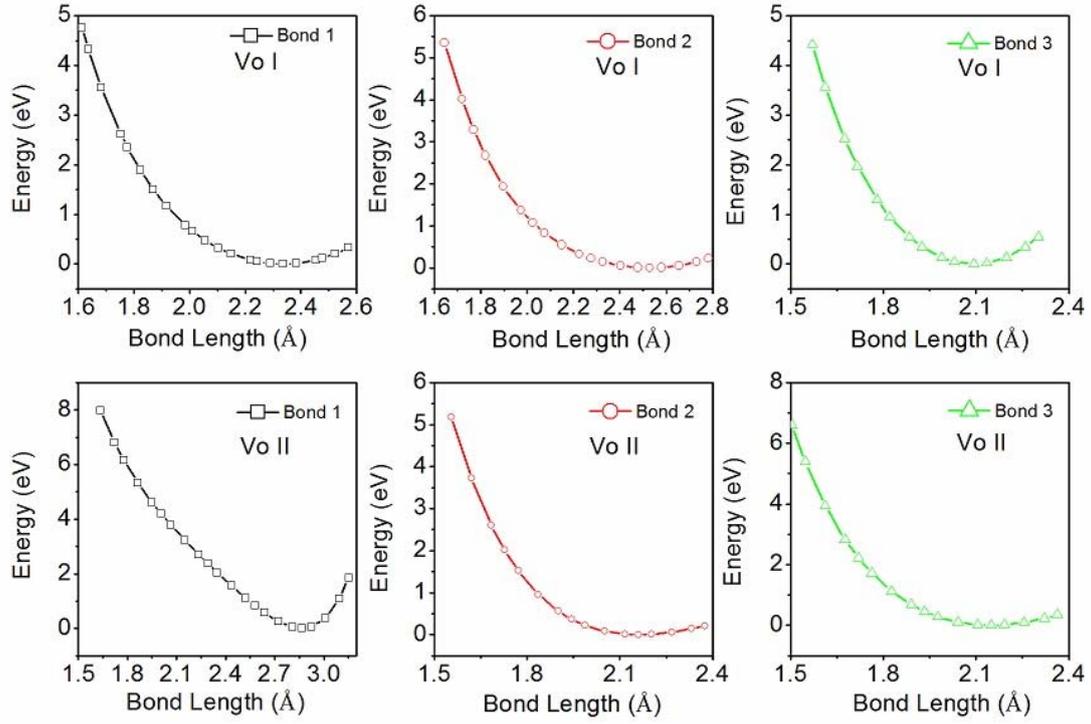

**FIG. 5** Relative energies of the oxygen-deficient $Ta_2O_5$ as a function of the bond lengths of the three Ta-O bonds marked in Fig. 1. The energies are obtained using static calculation, i.e., the positions of all the atoms are fixed for each calculation. The total energies of the referenced/relaxed configurations are set at 0.

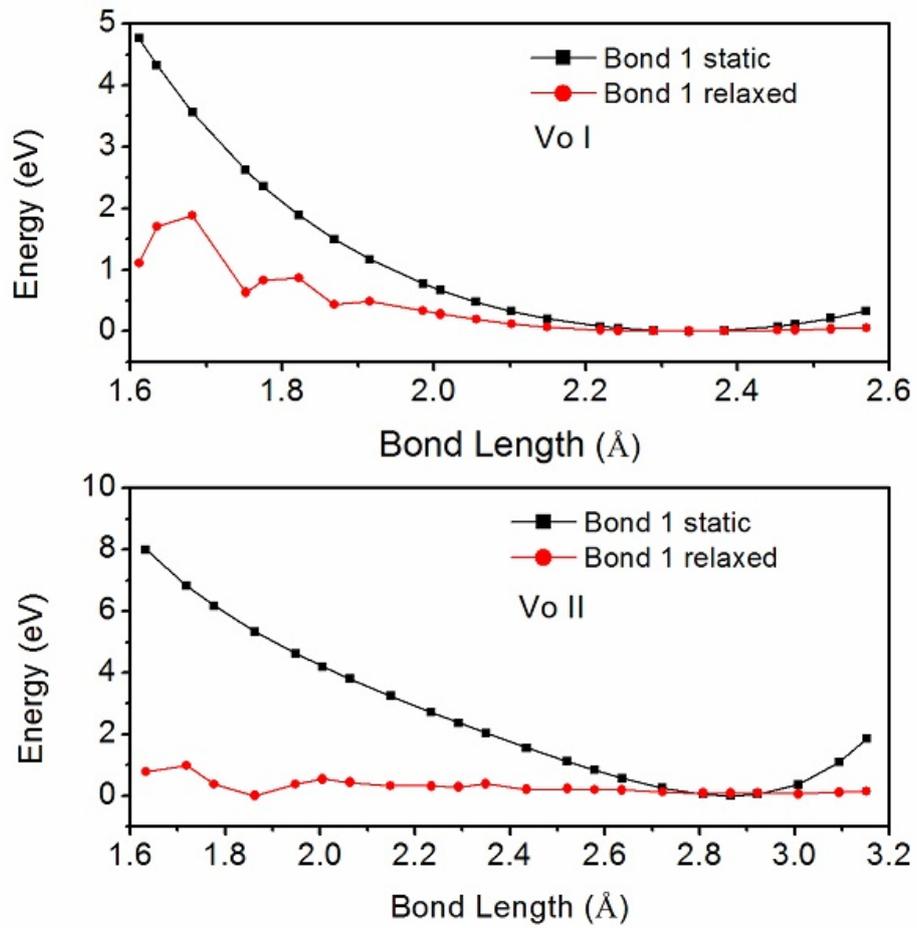

**FIG. 6** Total energy of the oxygen-deficient $Ta_2O_5$ as a function of the bond lengths of Bond 1 (Fig. 1), for the static and relaxed vacancy configurations Vo I and Vo II.